# Gender domain adaptation for automatic speech recognition

Artem Sokolov
Laboratory of Algorithms and Technologies for
Network Analysis
HSE University
Nizhny Novgorod, Russia
artsokol@hse.ru

Savchenko Andrey
Laboratory of Algorithms and Technologies for
Network Analysis
HSE University
Nizhny Novgorod, Russia
avsavchenko@hse.ru

*Abstract*—This paper is focused on the finetuning of acoustic models for speaker adaptation goal on a given gender. We pretrained the Transformer baseline model on Librispeech-960 and conduct experiments with finetuning on the gender-specific test subsets and. In general, we do not obtain essential WER reduction by finetuning techniques by this approach. We achieved up to ~5% lower word error rate on the male subset and 3% on female subset if the layers in the encoder and decoder are not frozen, but the tuning is started from the last checkpoints. Moreover, we adapted our base model on the full L2 Arctic dataset of accented speech and fine-tuned it for particular speakers and male and female genders separately. The models trained on the gen-der subsets obtained 1-2% higher accuracy when compared to the model tuned on the whole L2 Arctic dataset. Finally, we tested the concatenation of the pretrained x-vector voice embeddings and embeddings from conventional encoder, but its gain in accuracy is not significant.

*Keywords—automatic speech recognition, speaker adaptation, deep neural networks, gender-based acoustic models*

## I. INTRODUCTION

State-of-the-art speaker-independent automatic speech recognition systems (ASR) show very accurate results when they are trained on large datasets with various speaker samples. However, the typical challenge for system is the dependence of recognition quality on acoustic characteristics of particular speaker. Indeed, it is practically impossible to collect the training dataset that contains enough hours of voice examples for all age ranges, genders, and accents for a certain language. The adaptation to a voice of particular speaker is possible, but usually requires much data.

The issue could be relaxed if a model is adapted for the group of speakers as it typically done for recognition of accents, namely, by using additional training of a model on a small dataset of accented data, that makes it possible to significantly increasing the recognition quality [1].

If we consider acoustic model based on deep neural network, the same approach but for gender groups is not studied enough. Indeed, gender is an important speaker variability factor. Females and males obviously have differences in their voices. Con-sonant noise and vowel formants of female speakers tend to be located at higher frequencies. Phonation type also seems to depend on the speaker's gender. Female voices are often considered more breathy than male voices [2]. Significant differences were present between genders in the distribution of energy throughout the analyzed frequency values, taken from voice samples [3].

These facts could be considered to reduce the error rates separately for male or female gender group, if a gender of a speaker is assumed to be known. Acoustic model could be adapted by finetuning on the voice samples from only one gender. As a result, ASR engine will contain several acoustic models. An appropriate model is chosen depending on the gender or age of the speaker. Nowadays, gender recognition and age prediction tasks are actively studied either for video [4, 5] or audio modalities [6, 7].

Thus, in this paper we examined the tuning of acoustic model for the gender do-main. We used a neural network-based acoustic model and conduct experiments with two English datasets. Due to cross-gender voice differences are not equal for all languages [8] we choose datasets to cover native and not native speakers. The rest of the paper is organized as follows. Section II contains brief discussion of related literature. We discuss an approach to fine-tune acoustic models in Section III. Section IV contains experimental results for LibriSpeech and L2 Arctic datasets. Finally, concluding comments are given in Section V.

## II. RELATED WORK

Numerous studies have explored non-linguistic features differences between females and males and approaches to adapt various speech systems for groups based on gen-der, age, and accents [9, 10, 11]. Speaker adaptation techniques like maximum a-posteriori (MAP) and maximum likelihood linear regression (MLLR) are applied to ASR systems based on Gaussian mixture models (GMM) and Hidden Markov Models (HMM) [12, 13].

Acoustic models for gender domains could also be trained from scratch. The paper [14] reports accuracy improvement for gender-based GMM which was employed for parliament speech recognition. Authors used video modality to classify the domain gender by face and then choose the appropriate speaker depending on the acoustic model. That combination provides substantial reduction of the word error rate (WER).

The deep neural network (DNN) requires significantly much training data when compared to classical HMM-GMM (Gaussian Mixture Models) techniques. It means that often training from scratch is not possible way. There exist techniques for improvement of ASR quality for particular groups of speakers. The DNN-HMM model has been adapted for a group with specific age and gender group for Italian

The work is supported by RSF (Russian Science Foundation) grant 20-71-10010.

language [15]. The baseline general model was trained there using several datasets that included adult male, adult female, and children utterances. New acoustic models were obtained by the continuation of training with the same parameters on each target subset. Each adapted model outperforms the baseline on its target test set (adult male, adult female, children). Thus, the authors showed that pretrained acoustic model could be biased for a certain gender group.

Transfer learning was successfully approbated in [16] as an approach for adapting adult speech recognition systems for children. The next obvious way to train a neural network with lack of training data is finetuning. It starts new training for a new dataset but initialize the weights from pretrained network without substantial network modification. Ordinary is to involve a new optimizer, hyperparameters, and replace the output layer. This method of model adaptation was used for accented speech from non-native English speakers and people with amyotrophic lateral sclerosis [1]. The authors of the latter paper trained two base models with different architectures on a large dataset. In further, these pretrained models were employed to improve accuracy for people with a non-standard speech by model tuning on a smaller dataset collected for a targeted group of people with amyotrophic lateral sclerosis and for non-native speakers. This work showed that WER reduction on target small datasets could be obtained by neural network adaptation via finetuning.

One more popular approach to adapt models for certain speaker domain is the in-corporation of voice embeddings, e.g., i-vectors, x-vector, and d-vectors. The x-vector embeddings capture information and allow to build very accurate classifiers for speaker gender and speaker identification tasks [17] and outperforms i-vectors. The d-vectors showed its advantages for cocktail-party related tasks [18].

III. MATERIALS AND METHODS

*A. Data*

**LibriSpeech.** LibriSpeech is a common and popular dataset for speech recognition system experiments [23]. It includes transcribed ~960 hours of public domain audiobooks which are dictated by many speakers. Totally, for train purposes, the male sample of a LibriSpeech provides about 496 hours and the female is a bit above 465 hours. The validation part is about 8 hours as for males as females.

**L2 Arctic.** L2-Arctic corpus (v5.0) is a small dataset of non-native English speech [24]. It includes voice samples of 24 speakers (12 males and 12 females). One country dataset has 2 male and 2 female speakers. The total duration is 27.1 hours. We split the data into 90/10 and as in original paper [1]. We have chosen for validation dataset utterances without proper nouns.

*B. Acoustic Model*

In this paper, we finetune the pretrained neural network for gender-based domain. We choose the CTC-loss Transformer architecture, which provides competitive ASR results [19]. It is a sequence-to-sequence model, which is widely used in end-to-end speech processing. It consists of encoder and decoder parts and self-attention mechanism between. The encoder converts speech features into embeddings, self-attention learns the sequence information and provides it to the decoder. The last one could be considered as a small language model (LM), as it learns to predict characters or pieces according to language rules. The last experiments show that it could also show state-of-the-art-result as a part of other models [20].

We prepare Transformer base models for further experiments. The model was pre-trained with 960 hours of LibriSpeech dataset and we also prepare 2 models for each gender part of it. Thus, we have three base models for experiments.

The Librispeech-960 model was finetuned separately on both gender domains of LibriSpeech and L2 Arctic. We used ESPnet framework [21] that provides Pytorch implementation and pipeline for feature extraction inherited from Kaldi. Receipts for the LibriSpeech dataset are supported from-the-box. Feature extraction for L2 Arctic we added by ourselves.

In our experiments with embeddings, we used pretrained 512 dimensioned voice x-vectors from the LibriSpeech Text-to-Speech pipeline in the ESPNet receipt [22]. We also modified the ASR feature extraction pipeline and incorporate x-vectors into the training pipeline. Two approaches have been implemented, namely, x-vector could be added to embeddings came from the encoder or concatenated with them.

*C. Baseline*

We trained acoustic models during 100 epochs of LibriSpeech data. We used the default feature extraction pipeline for the dataset receipt in ESPnet. It assumes 80 filter banks as input for the acoustic model. To incorporate x-vectors we add MFCC (mel-frequency cepstral coefficients) extraction for each utterance. The model trained with a default for LibriSpeech receipt parameters: Noam optimizer, warmup step 25000, and transformer learning rate 5.

Class scores (estimates of posterior probabilities) at the output of Transformer acoustic model can be fused with separate RNN-based language model. But we assume that experiments without the external language model provides more fair comparison and do not use LM in our experiments.

*D. Finetuning*

We changed various hyperparameters, optimizers, number of epochs to achieve the best result. We saved checkpoints for all 100 epochs during training of acoustic mod-el, so that it is possible to start finetuning from any checkpoint. We conducted experiments with adaptation starting from the 20th, 50th, 60th, 80th, and 100th epochs. All experiments except beginning from 100th epochs finally lead to accuracy degradation. We tried to replace the Noam optimizer with Adadelta and Adam and play with its hyperparameters and the number of epochs. Our best result yielded with model tuning during 20 epochs with Adadelta, learning rate 0.1 and warmup step decreasing to 4000. All experiments were made. Our experiments with partial frozen layers con-firmed that Transformer finetuning provides the best result if we do not freeze any layers similarly to other Encoder-Decoder network LAS.

IV. EXPERIMENTATL RESULTS

*A. LibriSpeech*

We report detailed WER observing we measured for our three base models trained on LibriSpeech in Table I. The general baseline model shows lower WER on all test sets than trained on gender subsets only. The main reason here is the size of the training data. The general model has been trained on almost twice larger dataset when compared to the training on gender specific data only.

TABLE I. AVERAGE WER OF BASE MODELS, TRAINED ON LIBRISPEECH AND ITS GENDER SUBSAMPLES.

| Test set | Baseline (960h) | Male subset | Female subset |
|---|---|---|---|
| test_clean male | **4** | 7.4 | 8.7 |
| test_clean female | **4.6** | 9.8 | 7.9 |
| test_clean full | **4.3** | 8.6 | 8.3 |
| test_other male | **11.2** | 17.4 | 20.0 |
| test_other female | **9.3** | 18.4 | 14.7 |
| test_other full | **10.2** | 17.9 | 17.4 |
| dev_clean male | **4.4** | 8.9 | 9.5 |
| dev_clean female | **3.8** | 8.3 | 6.9 |
| dev_clean full | **4.1** | 8.6 | 8.2 |
| dev_other male | **11.2** | 17.0 | 20.5 |
| dev_other female | **9.3** | 18.8 | 14.6 |
| dev_other full | **10.2** | 17.8 | 17.5 |

TABLE II. AVERAGE WER OF FINETUNED MODEL ON MALE SUBSET OF LIBRISPEECH.

| Test set | Baseline (960h) | Male adapted |
|---|---|---|
| test_clean male | 4.0 | **3.8** |
| test_clean female | 4.6 | 4.6 |
| test_other male | 11.2 | **11.0** |
| test_other female | **9.3** | 9.4 |
| dev_clean male | 4.4 | **4.3** |
| dev_clean female | 3.8 | **3.7** |
| dev_other male | 11.2 | **11.0** |
| dev_other female | **9.3** | 9.5 |

TABLE III. AVERAGE WER OF FINETUNED MODEL ON FEMALE SUBSET OF LIBRISPEECH.

| Test set | Baseline (960h) | Female adapted |
|---|---|---|
| test_clean male | 4.0 | 4.0 |
| test_clean female | 4.6 | 4.6 |
| test_other male | **11.2** | 11.3 |
| test_other female | 9.3 | **9.1** |
| dev_clean male | 4.4 | **4.3** |
| dev_clean female | 3.8 | **3.7** |
| dev_other male | **11.2** | 11.4 |
| dev_other female | 9.3 | **9.1** |

The results of measurements for the finetuned model on the male subset of LibriSpeech is provided in Table II. We established that gender-based fine-tuning provides some non-significant accuracy improvement ~5%, but also it provokes the degradation on subsets for other gender group. Taken into account the results of [14] where the pretrained model continues to be learned on the part of the dataset we expected much more enhancement. We yielded up to 3% accuracy change with finetuning for female LibriSpeech subsample subset showed. These results are noted in Table III.

Thus, gender adaptation by the way of finetuning on subsamples base dataset allows achieving tiny error reduction only. This fact could be explained that our model is already well-trained and "overfit" on the same distribution is not able to bias the model significantly. When we started our tuning from an earlier checkpoint (when the model was not trained enough) we biased the model much more but did not achieve the best final result.

X-vector embeddings were examined in the next set of experiments. The training curve with the accuracy on the LibriSpeech validation dataset is shown in Fig. 1. Concatenation of x-vector and encoder embeddings gives results worse than the baseline. Hence, we conducted our measurement for the model, trained with summation of x-vectors. Table IV shows that x-vector incorporation does not permit to obtain considerable accuracy enhancement. Our adaptation experiments for the male subset does not show changes at all. Unfortunately, the model with voice embeddings is not suitable to enhance the overall quality of gender adaptation. Here our expectation that we could increase the difference in accuracy between baseline and finetuned model is not confirmed, so that x-vectors seem to be useless for this task.

TABLE IV. AVERAGE WER OF BASELINE TRAINED WITH X-VECTOR EMBEDDING.

| Test set | Baseline (960h) | Baseline with added X-vectors |
|---|---|---|
| test_clean | **4.3** | 4.4 |
| test_other | 10.2 | **10.1** |
| dev_clean | 4.1 | 4.1 |
| dev_other | 10.2 | 10.2 |

*B. L2 Arctic*

The next set of experiments is connected with L2 Arctic. We observed the changes of WER after tuning the base model on the test part of this dataset. Measurements of error for the L2 dev dataset and its gender subset are shown in Table V. We achieved similar results as in the paper from Google.

The final accuracy is improved on the accented dataset after finetuning. But our model showed less error reduction when compared to the RNN-Transducer [1]. Our model is not so powerful and we do not use the language model because we did not set the goal to achieve the maximum WER reduction.

We expected that cross-dataset finetuning on gender subsets should show more sensitive enhancement that it was in previous chapter. But our experimental results from Table VI show that adaptation on separated accented subsets yields again a non-significant profit 0.1-2%. As for the LibriSpeech dataset, gender female adaptation gives a bit low error than males.

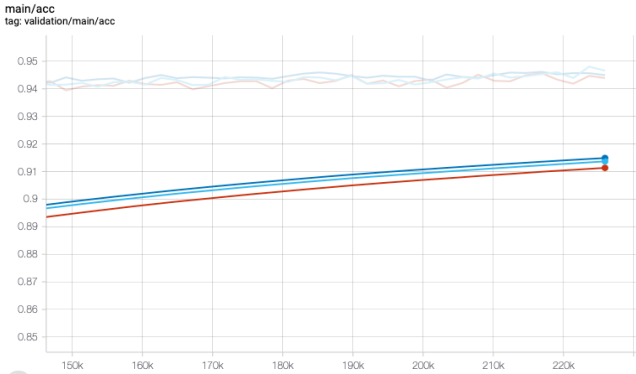

Fig. 1. Validation accuracy for training with voice embeddings: baseline (blue curve), baseline for summation of embeddings with x-vector (deep blue curve); Baseline for concatenation of embeddings with x-vector (red curve).

TABLE V.   AVERAGE WER OF CROSS-DATASET FINETUNED MODEL, L2 ARCTIC DATASET

| Test Set | Baseline (960h) | Finetuned (L2 Arctic train) |
|---|---|---|
| Arctic L2 dev full | 29.5 | **19.1** |
| Arctic L2 dev male | 27.7 | **20.1** |
| Arctic L2 dev female | 31.2 | **18.4** |

TABLE VI.   AVERAGE WER OF CROSS-DATASET FINETUNED MODEL, L2 ARCTIC DATASET

| Test Set | Arctic L2 train full | Arctic L2 train male | Arctic L2 train female |
|---|---|---|---|
| Arctic L2 dev male | 20.1 | **20.0** | - |
| Arctic L2 dev female | 18.4 | - | **18.1** |

## V. CONCLUSION

In this paper, we examined finetuning of DNN acoustic models to improve accuracy for male and female speaker by assuming that the gender of particular speaker is known. We trained the baseline Transformer model on LibriSpeech data and adapted it by male and female subsets of this dataset. It was experimentally found that better result gives common practice like learning rate and warmup step decreasing and changing of optimizer from Noam to Adadelta. We obtained a small error reduction (5%) when the baseline we bias on the male part of the dataset. However, incorporating x-vector between the encoder and decoder of the Transformer does not impact accuracy (Table IV). It leaves the same.

A series of experiments for L2 Arctic dataset with non-native English speech proved the considerably better accuracy as expected. This result is not significantly outperformed (0.1-2%) by cross-dataset adaptation on separate gender samples of L2 Arctic. In fact, we do not obtain essential WER reduction by finetuning techniques for both datasets. In future, it is necessary to train other acoustic features because default Mel frequency filter banks do not represent the vocal differences between females and males. We should try to add pitch features or replace filter banks with MFCCs. Moreover, it is necessary to repeat our experiments with other architecture because it could be more sensitive for finetuning. One more potential direction for adaptation we see is the meta-learning.